\documentclass[
reprint,
superscriptaddress,
amsmath,amssymb,
aps,
prb,
]{revtex4-2}

\usepackage{url}

\usepackage{xcolor}

\usepackage{graphicx}
\usepackage{dcolumn}
\usepackage{bm}
\usepackage{hyperref}

\usepackage{ulem}

\begin{document}

\preprint{APS/123-QED}

\title{Topological pump and its plateau transitions of $N$-leg spin ladder
}

\author{Kota Yamamoto}
\affiliation{
 Graduate School of Pure and Applied Sciences, University of Tsukuba, Tsukuba, Ibaraki 305-8571 \\
}

\author{Yoshihito Kuno}
\affiliation{
 Graduate School of Engineering Science, Akita University, Akita 010-8502\\
}
\author{Tomonari Mizoguchi}
\affiliation{
 Graduate School of Pure and Applied Sciences, University of Tsukuba, Tsukuba, Ibaraki 305-8571 \\
}
\affiliation{
 Department of Physics, University of Tsukuba, Tsukuba, Ibaraki 305-8571
}

\author{Kazuki Sone}
\affiliation{
 Graduate School of Pure and Applied Sciences, University of Tsukuba, Tsukuba, Ibaraki 305-8571 \\
}
\affiliation{
 Department of Physics, University of Tsukuba, Tsukuba, Ibaraki 305-8571
}
\author{Yasuhiro Hatsugai}
\affiliation{
 Graduate School of Pure and Applied Sciences, University of Tsukuba, Tsukuba, Ibaraki 305-8571 \\
}
\affiliation{
 Department of Physics, University of Tsukuba, Tsukuba, Ibaraki 305-8571
}

\date{\today}

\begin{abstract}
A topological pump on an $N\textrm{-}$leg spin ladder is discussed by introducing spatial clusterization whose adiabatic limit is a set of $2N\textrm{-}$site staircase clusters.
We set a pump path in the parameter space that connects two different symmetry protected topological phases.
By introducing a symmetry breaking staggered magnetic field, the system is always gapped during the pump.
In the topological pump {thus obtained}, the bulk Chern number is given by the number of the critical points enclosed by the pump path.
Plateau transitions characterized by the Chern number are demonstrated associated with deformation of the pump path. 
We find that there are $N$ critical points enclosed by the pump path for the $N\textrm{-}$leg ladder.
The ground state phase diagram without symmetry breaking terms is numerically investigated by using the quantized Berry phase.
We also discuss the physical picture of edge states in the diagonal boundary, and numerically demonstrate the bulk-edge correspondence for $N=2,3$ cases.
\end{abstract}

\maketitle

\section{\label{intro} introduction}
\par
The quantization of the charge transport for filled band electrons is discussed by Thouless \cite{thouless}.
This phenomenon is a $1+1$ dimensional analog of the quantum Hall effect as the Chern number characterizes the total amount of transported charge \cite{TKNN,thouless}.
It serves as the foundation of topological pumps, and other types of topological pumps have been also studied \cite{spin_pump_shindou,quasicrystal_pump}.
A few decades after Thouless's first proposal, the topological pump has attracted renewed interest due to its experimental realization using cold atoms \cite{lohse_2016_pump_experiment,nakajima_2016_pump_experiment} and classical simulators such as a photonic crystal \cite{Kraus2012,Jurgensen2021} and electric circuits \cite{Yatsugi2022,Stegmaier2024}.
In addition, a theoretical study on the topological pump for non-interacting fermions established its bulk-edge correspondence \cite{hatsugai_fukui}, which is a fundamental aspect of topological phenomena \cite{Hatsugai1993}.
Recently, there has been growing interest in extending topological pumps to interacting many-body systems {\cite{pump_strongly_interacting,pump_interacting_bRM,interaction_induced_pump,pump_eBH}}.
\par
Among varieties of topological pumps, a general idea is proposed in Ref.~\onlinecite{interaction_induced_pump} to construct a topological pump from a viewpoint of the symmetry protected topological (SPT) phases \cite{SPT2010,SPT2011,SPT2012} and the critical points. 
If the two different SPT phases with the same symmetries exist, then a gapless transition needs to exist in the symmetry-respecting parameter space. 
The critical point between the two is a source of the non-trivial topological pump. 
By setting a suitable parameter path enclosing the gapless point without gap closing, the pump can be topologically non-trivial since it is an obstruction for contracting the loop to a point.
\par
Gapped SPT phases are topologically stable against parameter deformation of the Hamiltonian unless its symmetry is broken.
Among various SPT phases, the Haldane phase~\cite{Haldane_gap,Haldane_gap2} is a prototypical and important one.
The Heisenberg spin chain of the integer spin is gapped according to the Haldane conjecture ~\cite{Haldane_gap,Haldane_gap2} that is supported by
various intensive studies, 
including analytical studies of the exactly solvable Affleck-Kennedy-Lieb-Tasaki model~\cite{AKLT}, and numerical studies~\cite{white_haldane_gap, QMC}.
\par
A spin ladder has been also examined in connection with the Haldane conjecture.
It is gapped(critical) when the number of legs is even(odd) \cite{Sierra_ladder,even_odd_conj_1996,even_odd_conj_1997,white_even_odd}, (even-odd conjecture).
Moreover, the spin ladder with various dimerization exhibits a phase diagram whose critical lines have been also investigated \cite{Sierra_dimerized_ladder,Sierra_dimerized_ladder_dmrg}.
In studying the SPT phases of the quantum spin systems including the spin ladders, the quantized Berry phase is useful as a topological order parameter under the time-reversal symmetry~\cite{hatsugai_berry_2006,hatsugai_berry_2007,hatsugai_berry_2010_spt,berry_ladder_general_spin_Mila,Fubasami}. 
\par
In this paper, we construct and analyze a model of a topological pump of an $N$-leg spin ladder by extending the idea of Ref.~\onlinecite{interaction_induced_pump}.
The spin ladder has a rich phase structure as the SPT phases, leading to several gap-closing points or lines between the different SPT phases. 
It implies the existence of various topological pumps. 
Thus, the $N$-leg spin ladder is an interesting playground to realize rich topological phenomena. 
We clarify the presence of SPT phases and characterize these phases by using the Berry phase.
The gauge structure of the spin ladders is also discussed.
The gap closing points as the topological obstruction are used to construct a topologically non-trivial pump. 
The plateau transitions of the topological pump are also demonstrated by varying the form of the pump path, characterized by the bulk Chern number. 
We further elucidate the boundary physics of the topological pump under open boundaries. 
The diagonal shape of the boundary is used for a clear understanding of the edge states and the spin center of mass (sCoM).
We numerically investigate the behavior of the sCoM in detail. 
Furthermore, we verify that it is related to the bulk Chern number, that is, the bulk-edge correspondence of the topological pump of the $N$-leg ladder is satisfied.
\par
The paper is organized as follows.
In Sec.~\ref{model}, we introduce an $N$-leg ladder with spatial modulation whose adiabatic limit consists of a set of $2N$-site staircase clusters, and investigate its ground state phase diagram.
The $2N\textrm{-}$site clusterization for the $N$-leg ladder is conceptually used as an extension of dimerization in spin chains \cite{plateau_transition,SUQ}.
We see that the $2N\textrm{-}$site clusterized $N$-leg ladder experiences $N$-times SPT phase transitions between two opposite modulation limits.
Furthermore, to characterize the phase transition, we employ the quantized Berry phase and show the results of its numerical calculation.
In Sec.~\ref{pump_in_PBC}, we construct a topological pump by introducing a staggered magnetic field to the $N$-leg ladder. 
We set pump parameter paths based on the phase diagram obtained in Sec.~\ref{model}. 
Numerical calculation of the Chern numbers that characterize the topological pumps and plateau transitions are demonstrated.
In Sec.~\ref{pump_in_OBC}, we discuss a topological pump with the open boundary condition by employing the DMRG method.
We also provide physical pictures of the edge states of the ladder in connection to the bulk-edge correspondence.
Sec.~\ref{summary} is a summary of this paper.

\section{\label{model} model}
We first introduce the model and discuss its properties of the gauge transformations. 
Then, the phase transitions between the SPT phases are numerically confirmed by employing the Berry phase, an efficient tool to detect transition points even for small-system-size numerics \cite{Kariyado2015}.

\subsection{ $N$-leg ladder and $2N$-site clusterization}
\par
\begin{figure}[t]
\includegraphics[keepaspectratio,scale=0.37]{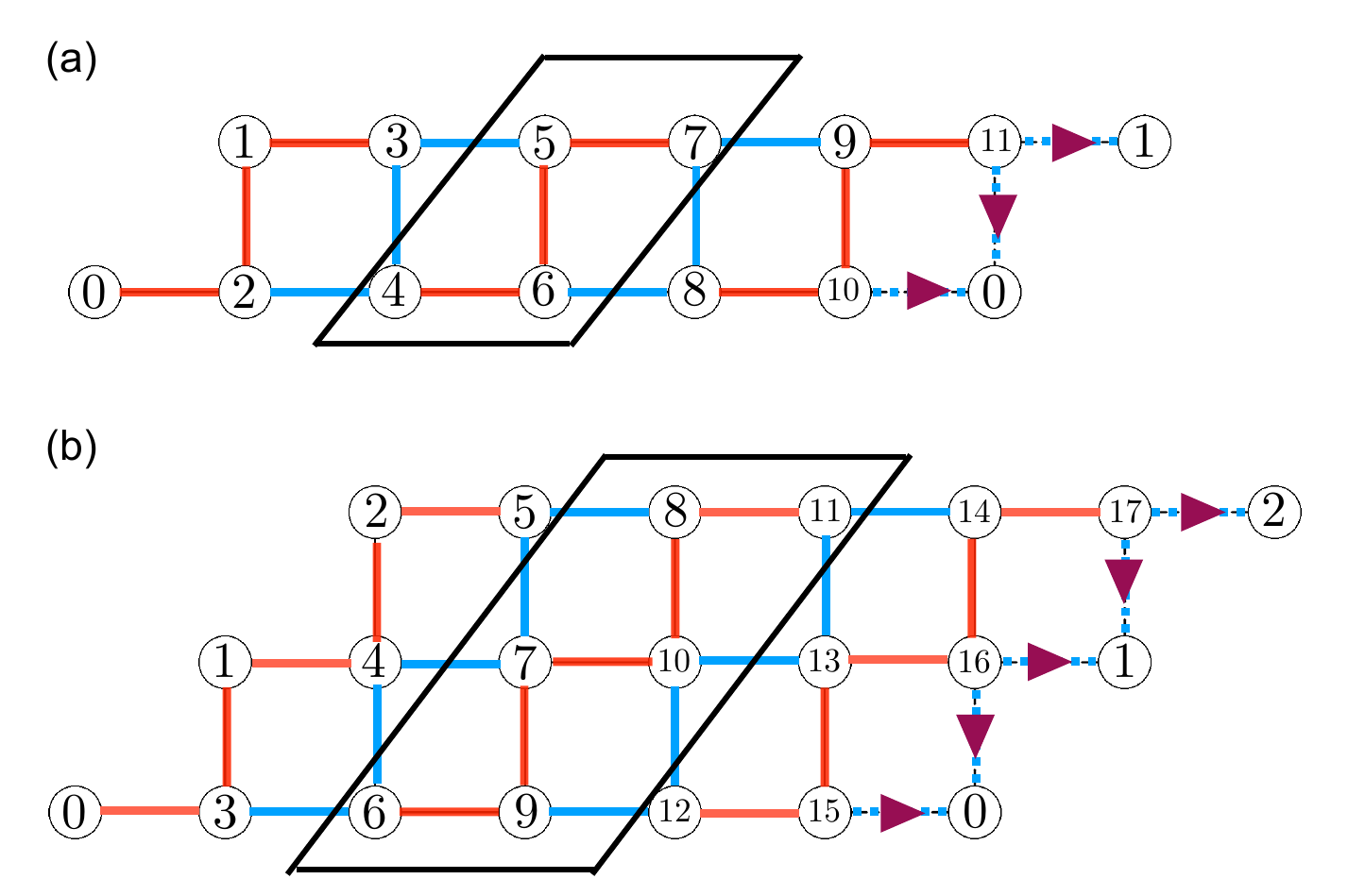}
\caption{Schematic of the system and the modulation pattern of the exchange couplings
for (a) $N=2$, $L=12$ and (b) $N=3$, $L=18$. The red bond represents $J-\delta$, while the blue bond represents $J+\delta$. The sites enclosed by the parallelogram represent the unit cell. The colored solid arrows represent the twist $e^{i\theta}$.
}
\label{fig:clusterization} 
\end{figure}

We start with a spin-half $N\textrm{-}$leg ladder. The Hamiltonian is
\begin{eqnarray}
    H &&= H_{\mathrm{leg}}+H_{\mathrm{rung}}, 
    \\
    H_{\mathrm{leg}} &&= \sum_{n=0}^{N-1}\sum_{j=0}^{\frac{L}{N}-1} J_{n+Nj} \mathbf{S}_{n+Nj} \cdot \mathbf{S}_{n+N(j+1)}, 
    \\
    H_{\mathrm{rung}} &&= \sum_{n=1}^{N-1}\sum_{j=0}^{\frac{L}{N}-1} K_{n+Nj} \mathbf{S}_{n+Nj} \cdot \mathbf{S}_{n+N(j+1)-1}, 
\end{eqnarray}
where $\mathbf{S}_{j} = (S_j^x,S_j^y,S_j^z)$ is the spin operator at the site $j$ ($j=0,\cdots,L-1$), the total number of sites is denoted by $L$, and the number of sites per leg $L/N$ is assumed to be an even integer.
For the index larger than $L-1$, it is replaced by its difference from $L$, i.e., $\mathbf{S}_{j+L}=\mathbf{S}_{j}$, i.e., we impose the periodic boundary condition.
It is also assumed that the exchange couplings are antiferromagnetic, $J_{n+Nj}(>0)$ and $K_{n+Nj}(>0)$. 
The schematic of the system is shown in Fig.~\ref{fig:clusterization}.
\par
We set the exchange couplings $J_{n+Nj}$ and $K_{n+Nj}$ to be clusterized as
\begin{eqnarray}
J_{n+Nj} = \left\{
\begin{array}{ll}
J - \delta & \mathrm{for} \> j \in 0,...,L/N-2 \\
J + \delta  & \mathrm{for} \> j \in 1,...,L/N-1 
\end{array}
\right.,
\\
K_{n+Nj} = \left\{
\begin{array}{ll}
J - \delta & \mathrm{for} \> j \in 0,...,L/N-2 \\
J + \delta & \mathrm{for} \> j \in 1,...,L/N-1 
\end{array}
\right..
\end{eqnarray}
Unless otherwise noted, we set $J=1$ and adopt it as a unit of the energy.
The modulation of the exchange couplings is illustrated in Fig.~\ref{fig:clusterization} for $N=2,3$.
Note that a unit cell consists of $2N$ sites, which are outlined by the parallelogram in Fig.~\ref{fig:clusterization}.

\subsection{ Twisted boundary condition}
\par
We now introduce the twisted boundary condition to define the Berry phase and the Chern number.
The twisted boundary is introduced by performing the large gauge transformation of a Hamiltonian with a global twist whose derivative with respect to the twist is associated with a spin current of the topological pump \cite{hatsugai_fukui,plateau_transition}.
\par
\begin{figure}[t]
\includegraphics[keepaspectratio,scale=0.38]{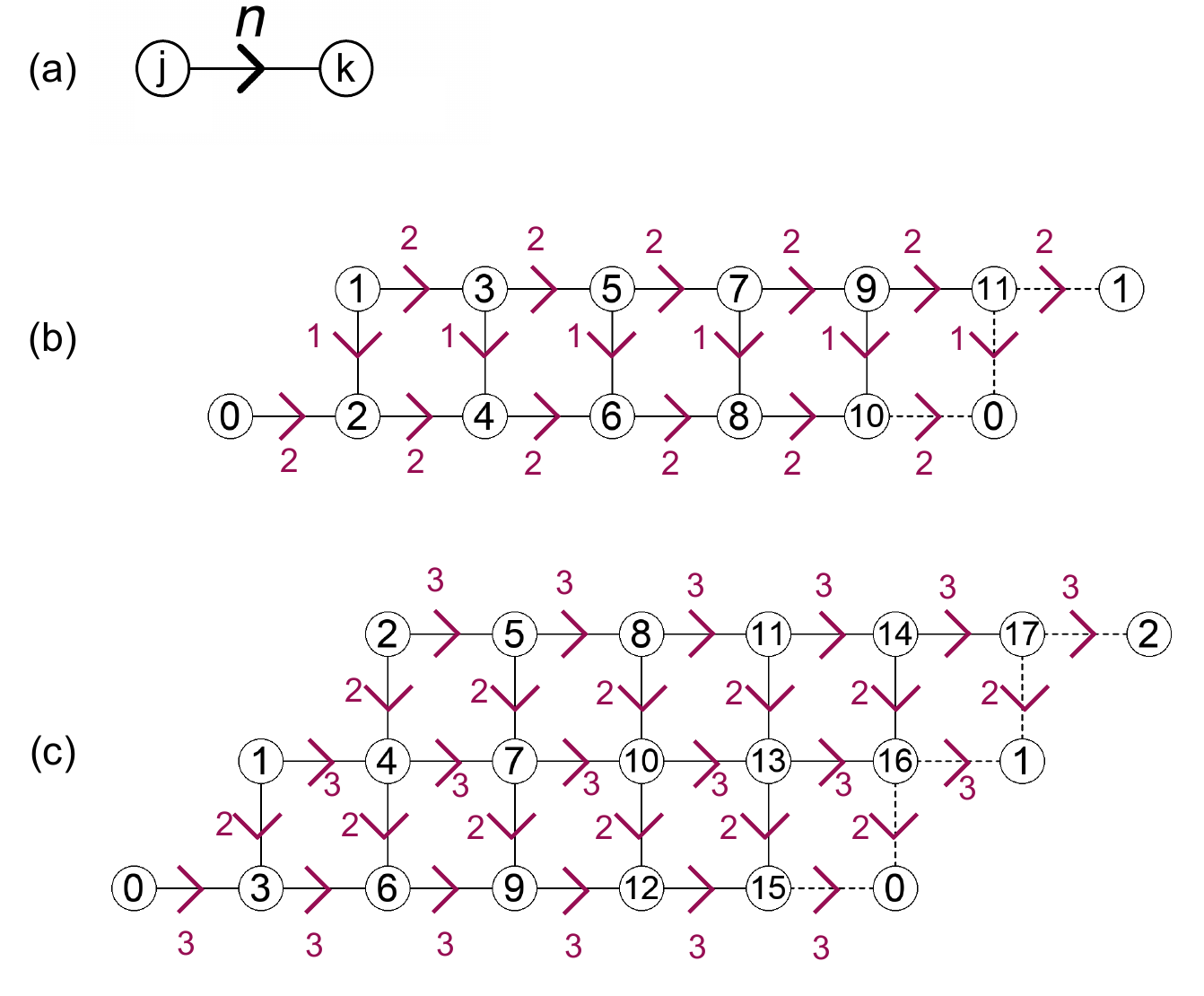}
\caption{ (a) The twist on the bond $\langle{j,k}\rangle$ where the coupling is modified as $\mathbf{S}_{j}\cdot\mathbf{S}_{k} \rightarrow  S^{z}_{j} S^{z}_{k} +\frac{1}{2} ( e^{i\theta\frac{n}{L}} S^{+}_{j} S^{-}_{k} + \mathrm{H.c.})$. The modified Hamiltonian with the global twist for (b) $N=2$, $L=12$ and (c) $N=3$, $L=18$.
}
\label{fig:global_twist} 
\end{figure}

The Hamiltonian of the ladder with the global twist, represented by $\bar{H}(\theta)$, is introduced as
\begin{eqnarray}
    \bar{H}(\theta) =&& \sum_{n=0}^{N-1}\sum_{j=0}^{\frac{L}{N}-1} J_{n+Nj} h_{n+Nj,n+{N(j+1)}} \left(\frac{N}{L}\theta \right) \nonumber\\&&+
    \sum_{n=1}^{N-1}\sum_{j=0}^{\frac{L}{N}-1} K_{n+Nj} h_{n+Nj,n+{N(j+1)}-1}\left(\frac{N-1}{L}\theta \right),\nonumber\\
\end{eqnarray}
where $h_{j,k}(\phi)$ is the $\phi\textrm{-twisted}$ exchange coupling between $\mathbf{S}_{j}$ and $\mathbf{S}_{k}$ as
\begin{eqnarray}
    h_{j,k}(\phi) = S^{z}_{j} S^{z}_{k} +\frac{1}{2} ( e^{i\phi} S^{+}_{j} S^{-}_{k} + e^{-i\phi} S^{-}_{j} S^{+}_{k}).\nonumber
\end{eqnarray}
The global pattern of the twists of $\bar{H}(\theta)$ is illustrated in Fig.~\ref{fig:global_twist}
\par
For the Hamiltonian $\bar{H}(\theta)$, we then apply the large gauge transformation as
\begin{eqnarray}
    U(\theta) = \prod_{j=0}^{L-1} u_j(\theta), \label{eq:large_gauge_transformation}\\
    u_j(\theta)=e^{-i\theta\frac{j-j_{\mathrm{c}}}{L} S_j^z},
\end{eqnarray}
where $j-j_{\mathrm{c}}$ is the coordinate measured from the center of the system, $j_{\mathrm{c}}=\frac{L-1}{2}$.
From this gauge transformation, the Hamiltonian with the local boundary twist, which we mainly focus on, is given by
\begin{eqnarray}
    H(\theta) = U^{\dagger}(\theta) \bar{H}(\theta) U(\theta).
\end{eqnarray}
The twisted boundary condition of the Hamiltonian $H(\theta)$
is illustrated in Fig.~\ref{fig:clusterization} by colored solid arrows.

\subsection{ Berry phase }
\par
We study the ground state phase diagram of the Hamiltonian $H(\theta)$ for the clusterization amplitude $\delta$.
The SPT phases are characterized by using the Berry phase quantized as $0,\pi$ under the time-reversal symmetry~\cite{hatsugai_berry_2006,hatsugai_berry_2007,hatsugai_berry_2010_spt}.
The Berry phase is given as 
\begin{eqnarray}
    \gamma = -i \int_{0}^{2\pi} d\theta \langle g(\theta) | \partial_{\theta} g(\theta) \rangle,
\end{eqnarray}
where $|g(\theta)\rangle$ is the ground state of $H(\theta)$, which we assume to be unique and gapped.
\par
\begin{figure}[t]
\includegraphics[keepaspectratio,scale=0.45]{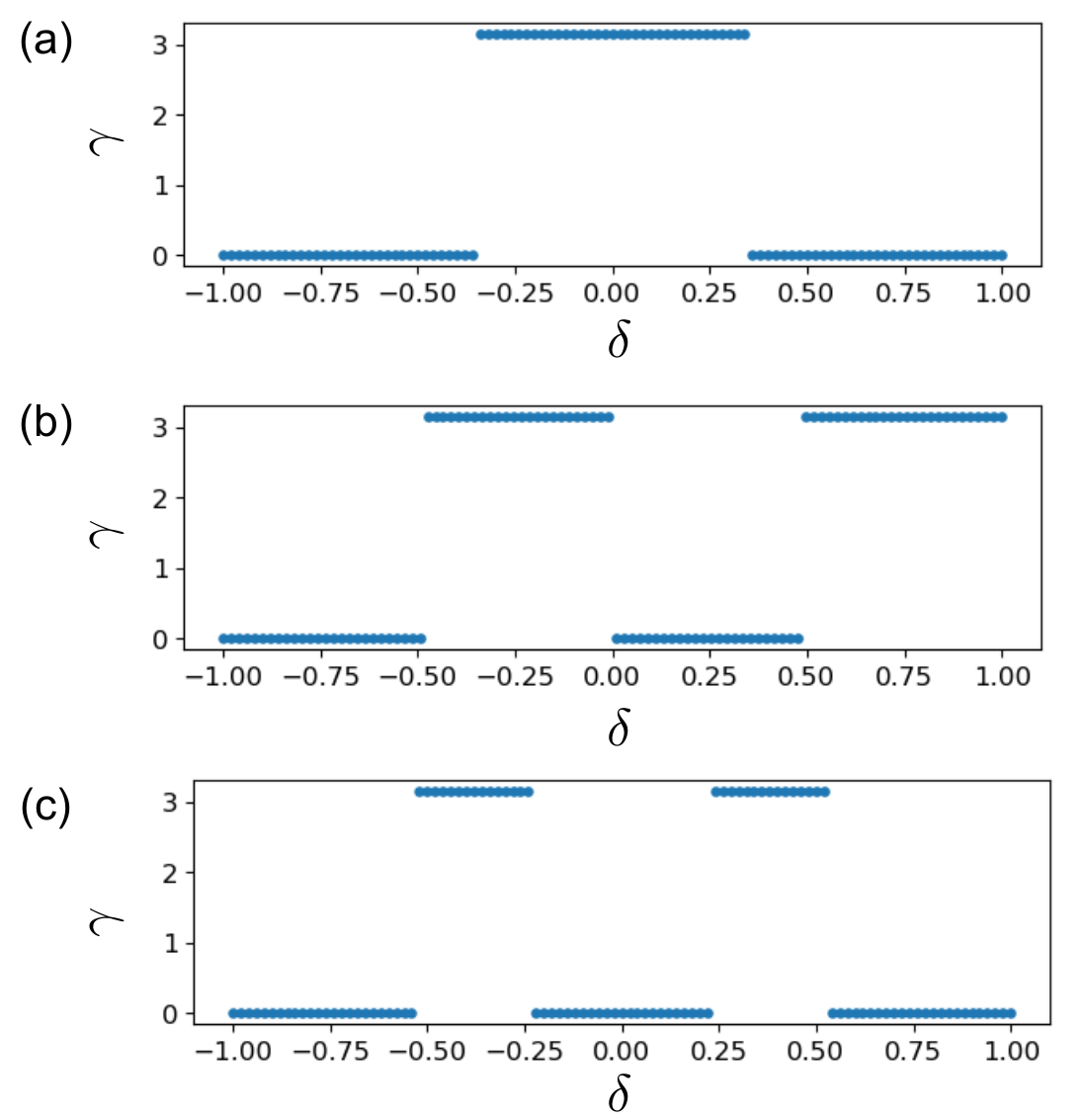}
\caption{The quantized Berry phase as a function of the modulation amplitude $\delta$ for (a) $N=2$, $L=16$, (b) $N=3$, $L=18$, and (c) $N=4$, $L=16$.
}
\label{fig:berry} 
\end{figure}

We first numerically calculate the $\delta$-dependence of the Berry phase.
The results for $N=2,3,4$ are shown in Fig.~\ref{fig:berry}. 
The exact diagonalization is used within the sector of $S^{z}=\sum_{j=0}^{L-1}S_{j}^{z}=0$. 
To calculate the Berry phase, we have used the discretization method of Fukui-Hatsugai-Suzuki~\cite{fukui-hatsugai-suzuki,hatsugai_berry_2006}. 
For each case, discontinuous changes of the Berry phase are obtained, indicating the existence of critical points where the ground state is degenerated at a certain value of $\theta$ for the finite system size. 
\par
The behavior of the Berry phase and the number of the critical points can be understood from the view of a local gauge transformation~\cite{degenaracy_berry}.
We consider a single valued local gauge transformation $V(\theta)$ given by
\begin{eqnarray}
V(\theta)=\prod^{L-1}_{j=L-N}v_j(\theta),\;\: v_j(\theta) = e^{i\theta(S_{j}^{z}-S)},
\end{eqnarray}
where $S = 1/2$.
Here, we rewrite $H(\theta)\to H(\theta,\delta)$ to represent $\delta$ dependence explicitly.
By applying this local on-site gauge transformation to $H(\theta)$, the twisted bonds of $H(\theta,\delta)$ are shifted to the left. 
This means that the sign of $\delta$-dependence is changed as
\begin{eqnarray}
    V^{\dagger}(\theta) H(\theta,\delta) V(\theta) = H(\theta,-\delta).
    \label{gt_relation}
\end{eqnarray}
In fact, the Berry phase for $V^{\dagger}(\theta) H(\theta,\delta) V(\theta)$ is $\gamma(\delta)+N\pi$~\cite{cal_BP}
where $\gamma(\delta)$ is the one for $H(\theta)$. 
Also, it is the Berry phase for the Hamiltonian $H(\theta,-\delta)$.
Then, we have
\begin{eqnarray}
    {\gamma(+\delta)+N\pi = \gamma(-\delta)} \quad(\mathrm{mod}\,2\pi).
\label{even_odd_function}
\end{eqnarray}
Thus, the Berry phase $\gamma(\delta)$ is an even (odd) function of $\delta$ when $N$ is even (odd) as long as the Berry phase is well-defined.
Especially, for the odd $N$ case with $\delta=0$, Eq.~(\ref{even_odd_function}) becomes $\gamma(0)+\pi=\gamma(0)$, which is a contradiction.
It implies the Berry phase is not well-defined, that is, the gap closes at some $\theta$.
Note that similar arguments based on the translational (or inversion) symmetry were discussed in Ref.~\cite{degenaracy_berry}.
\par
To further obtain physical insights, we discuss two extreme cases:
\par
\underline{\bf Case $\delta=J$}: 
In this case, the blue bonds in Fig.~\ref{fig:clusterization} are dominant, 
and thus the Berry phase is determined by the $2N$-site cluster with the twist, which is equivalent to the open $2N$-site chain.
The Berry phase of this $2N$-site chain is calculated explicitly.
To be concrete, we consider a local gauge transformation which is single valued as $v_j(\theta)\equiv v_j(\theta+2\pi)$ given by
\begin{eqnarray}
    v_j(\theta) = e^{i\theta(S_{j}^{z}-S)},
\end{eqnarray}
where $S = 1/2$~\cite{degenaracy_berry}.
The decoupled twisted {$2N$-site} chain is obtained by performing $\prod_{j=0}^{N-1} v_j(\theta)$ to the untwisted open chain.
For a local cluster state without twist, $|\rho\rangle$, the local cluster with finite twist is given by applying $\prod_{j=0}^{N-1} v_j(\theta)$ to $|\rho\rangle$. 
We then consider the Berry phase for the state $\prod_{j=0}^{N-1} v_j(\theta)|\rho\rangle$ as $\gamma_{\mathrm{cluster}}=-i\int^{2\pi}_{0}(i NS)d\theta =N\pi \;\;\quad(\mathrm{mod}\,2\pi)$.
Thus, the Berry phase of the decoupled twisted chain with $2N$ sites is $N\pi$.
\par
\underline{\bf Case $\delta=0$}: The ladder has the translational symmetry whose unit cell consists of $N$ sites.
As shown in Fig.~\ref{fig:berry}(a), the Berry phase for $\delta=0$ and $N=2$ becomes $\pi$, which is consistent to the previous studies~\cite{berry_ladder_general_spin_Mila,degenaracy_berry}.
The ground state phase is the rung singlet phase, which is the product state of decoupled rung dimers, and its physical properties are investigated in various approaches \cite{rung_singlet_spectrum,rung_singlet_white_dmrg,rung_singlet_magnon}, including the Berry phase analysis \cite{ring_exchange,berry_ladder_general_spin_Mila}.
The above arguments on the two limiting cases are all consistent with our numerical results in Fig.~\ref{fig:berry}.
\section{\label{pump_in_PBC} Pump with periodic boundary}
\par
In the previous section, we demonstrate rich phase diagrams of the spin ladder with different SPT phases and gapless transition points.
The transition points separating the different SPT phases are sources of the non-trivial topological pump. 
In this section, we construct the topological pump with the periodic boundary condition based on the clusterized spin ladder. Then, we verify the topological pump of the bulk by numerically calculating the Chern number.
\par
Firstly, a staggered magnetic field is introduced, and a pump Hamiltonian is constructed by setting periodic parameters in time.
The additional Hamiltonian is given as
\begin{eqnarray}
    H_{\mathrm{SM}}(t) &&= \sum_{j=0}^{L-1} f_{j} \Delta(t) {S}_{j}^{z}, 
    \\
    \Delta(t) &&= \Delta_{0} \sin \left(2\pi\frac{t}{T} \right),
\end{eqnarray}
with
\begin{eqnarray}
f_j = \left\{
\begin{array}{ll}
-1 & j \equiv {0,...,N-1\,(\mathrm{mod}\,2N)} \\
+1 & j \equiv {N,...,2N-1\,(\mathrm{mod}\,2N)} 
\end{array}
\right..
\end{eqnarray}
Here, $\Delta_0$ is the strength of the field and $T$ is the time period of the pump.
The term $H_{\mathrm{SM}}(t)$ breaks the time-reversal symmetry except for $t=0$ and $T/2$. 
It is a symmetry breaking term for the SPT phases discussed in the previous section.
\par
We also make the exchange parameters time-dependent as $H(\theta,t)=H(\theta,t+T)$.
The couplings $J_{n+Nj}$ and $K_{n+Nj}$ are set as
\begin{eqnarray}
J_{n+Nj}(t) &&= \left\{
\begin{array}{ll}
J - \delta(t) & j \in \mathrm{even} \\
J + \delta(t) & j \in \mathrm{odd} 
\end{array}
\right.,\nonumber\\
K_{n+Nj} (t) &&= \left\{
\begin{array}{ll}
J - \delta(t) & j \in {\mathrm{even}} \\
J + \delta(t) & j \in {\mathrm{odd}} 
\end{array}
\right.,
\\
\delta(t) &&= \delta_{0}  \cos\left(2\pi\frac{t}{T} \right).
\end{eqnarray}
The pump Hamiltonian is then written as
\begin{eqnarray}
    H_{\mathrm{P}}(\theta,t) = H(\theta,t)+H_{\mathrm{SM}}(t). \label{eq:pump_Hamiltonian}
\end{eqnarray}
The path of the pump parameters $\delta(t)$ and $\Delta(t)$ are set to enclose the critical points on the $\Delta=0$
line in the parameter space with $\delta$ and $\Delta$. 
Here, if the ground state of the pump Hamiltonian is unique for any value of $\theta$ and $t$ \cite{interaction_induced_pump,plateau_transition}, the topological pump can appear.
\par
Based on this setup, we characterize the pump by using the Chern number \cite{thouless,haldane_spin_chern}, which is equal to the pumped spin per cycle as the time integral of the spin current \cite{hatsugai_fukui,plateau_transition} (see Appendix).
The Chern number is given as
\begin{eqnarray}
C=  \frac{1}{2\pi i} \int_0^{2\pi} d\theta \int_0^T dt (\partial_{\theta} A_{t}(\theta,t) - \partial_{t} A_{\theta}(\theta,t)),
\end{eqnarray}
where $A_{\mathrm{\alpha}}(\theta,t) = \langle g(\theta,t) | \partial_{\alpha} g(\theta,t) \rangle$ ($\alpha=\theta, t$) is the Berry connection in the $\theta\textrm{-}t$ space with $|g(\theta,t) \rangle$ being the unique ground state of the snap-shot Hamiltonian $H_{\mathrm{P}}(\theta,t)$.
\par
We numerically calculate the Chern number by using the exact diagonalization and the Fukui-Hatsugai-Suzuki formula \cite{fukui-hatsugai-suzuki}.
The system size number $L$ is set to be even and the amplitude of the magnetic field $\Delta_0$ is kept weak enough~\cite{plateau_transition}. 
\par
\begin{figure}[t]
\includegraphics[keepaspectratio,scale=0.55]{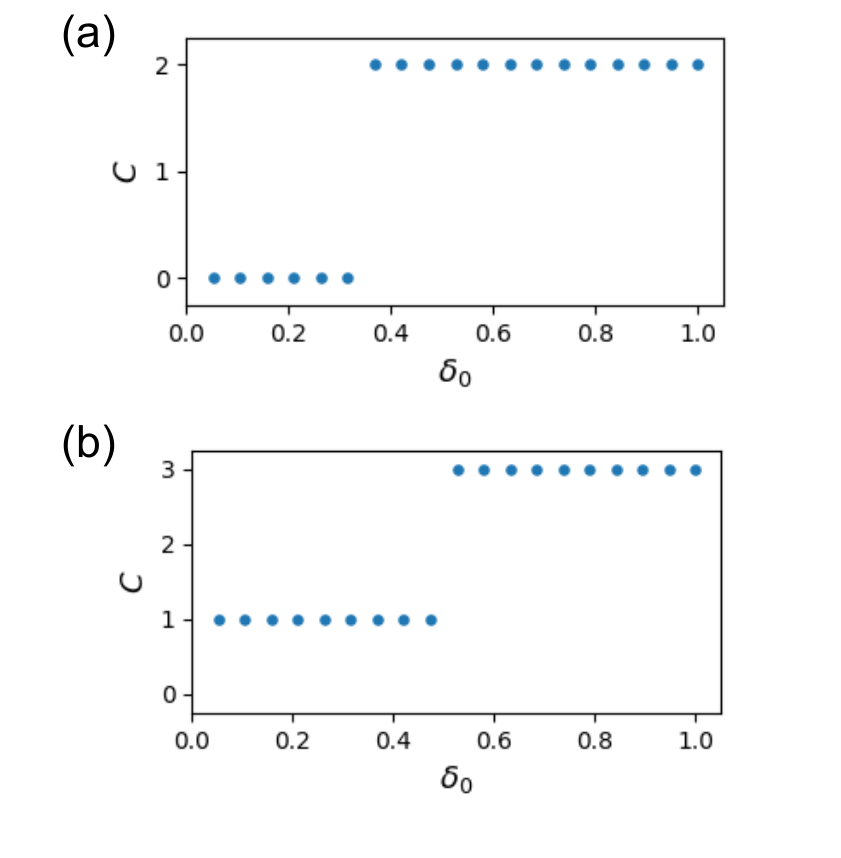}
\caption{The Chern number for the modulation amplitude $\delta_0$ for (a) $N=2$, $L=16$ and (b) $N=3$, $L=18$. We set $\Delta_0=0.1$. 
}
\label{fig:chern} 
\end{figure}

We calculate the Chern number as varying $\delta_0$ corresponding to the radius of the pump path.
The results for $N=2$, $3$ are shown in Fig.~\ref{fig:chern}.
The maximum Chern number we obtained is $C=N$ for $\delta_0=J$, which is equivalent to the number of critical points
on the $\Delta=0$ line in the $\delta\textrm{-}\Delta$ parameter space.
We also confirm the plateau transitions due to the Chern number~\cite{plateau_transition} when changing the modulation amplitude $\delta_0$ in Fig.~\ref{fig:chern}.
The Chern number changes only if the pump path crosses the critical gapless points.
These behaviors are reminiscent of the plateau transitions that have been studied in the quantum Hall systems\cite{qhs_plateau1999_4,qhs_plateau1999_9,lattice_fermion_plateau,plateau_graphene}.

\section{\label{pump_in_OBC} Pump with open boundary}
In the previous sections, we demonstrate the topological pump of the bulk.
Here we consider the system with open boundary conditions to demonstrate the bulk-edge correspondence of the topological pump.
Namely, the role of edge states in the topological pump is elucidated.
\par
To argue the bulk-edge correspondence, we define a spin center of mass (sCoM):
\begin{eqnarray}
P(t)=\sum_{j=0}^{L-1}\langle g(t)|\frac{j-j_{\mathrm{c}}}{L}{S}^z_{j}|g(t)\rangle,
\end{eqnarray}
where $|g(t)\rangle$ is the unique ground state of the topological pump Hamiltonian with the open boundary condition (given later). 
The time derivative of $P(t)$ gives a spin current, but $P(t)$ itself can be discontinuous during the pumping due to the degeneracy associated with the edge states (See Appendix).
We denote $t_i$ as the time when the sCoM $P(t)$ jumps discontinuously.
\par
The bulk-edge correspondence of the topological pump is 
written as~\cite{hatsugai_fukui,interaction_induced_pump,pump_eBH,plateau_transition},
\begin{eqnarray}
I = C,\nonumber
\end{eqnarray}
where $I$ is the sum of the discontinuous jumps of the sCoM per cycle written as
\begin{eqnarray}
I = -\sum_{i} {P}(t_i)|_{t_i-0}^{t_i+0}.
\end{eqnarray}
See Appendix~\cite{hatsugai_fukui,plateau_transition}.
Note that $I$ is also quantized in the large system size limit $L\rightarrow\infty$.
\par
\begin{figure}[t]
\includegraphics[keepaspectratio,scale=0.33]{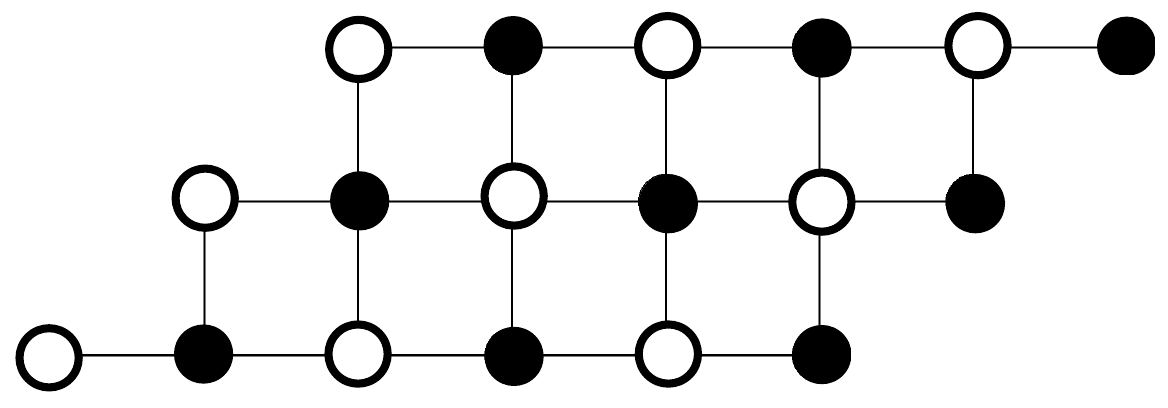}
\caption{Diagonal boundary and its boundary sites:
Sites in a sublattice are depicted by circles, and sites in the other sublattice are depicted by black filled circles. Spins at each end of the ladder are in the same sublattice.
}
\label{fig:edge} 
\end{figure}

We define the pump Hamiltonian with the open boundary condition by cutting the boundary bonds in the periodic pump Hamiltonian, keeping the other pump parameters unchanged.
To be concrete, the pump Hamiltonian is given as
\begin{eqnarray}
    H_{\mathrm{O}}(t) &&= \sum_{n=0}^{N-1}\sum_{j=0}^{\frac{L}{N}-2} J_{n+Nj}(t) \mathbf{S}_{n+Nj} \cdot \mathbf{S}_{n+Nj+N} \nonumber\\
    &&+ \sum_{n=1}^{N-1}\sum_{j=0}^{\frac{L}{N}-2} K_{n+Nj}(t) \mathbf{S}_{n+Nj} \cdot \mathbf{S}_{n+Nj+N-1} \nonumber\\
    &&+H_{\mathrm{SM}}(t). 
    \label{eq:H_o}
\end{eqnarray}
We here introduce the diagonal edge boundary shown in Fig.~\ref{fig:edge}.
By considering the diagonal edge, the physical picture of the edge states is simplified since the direction of the onsite magnetic field is the same for all the spins at each end of the ladder.
It is known that two-leg ladders have edge states when the shape of the edges is diagonal \cite{ring_exchange,ring_exchange_2}.
\par
\begin{figure}[t]
\includegraphics[keepaspectratio,scale=0.33]{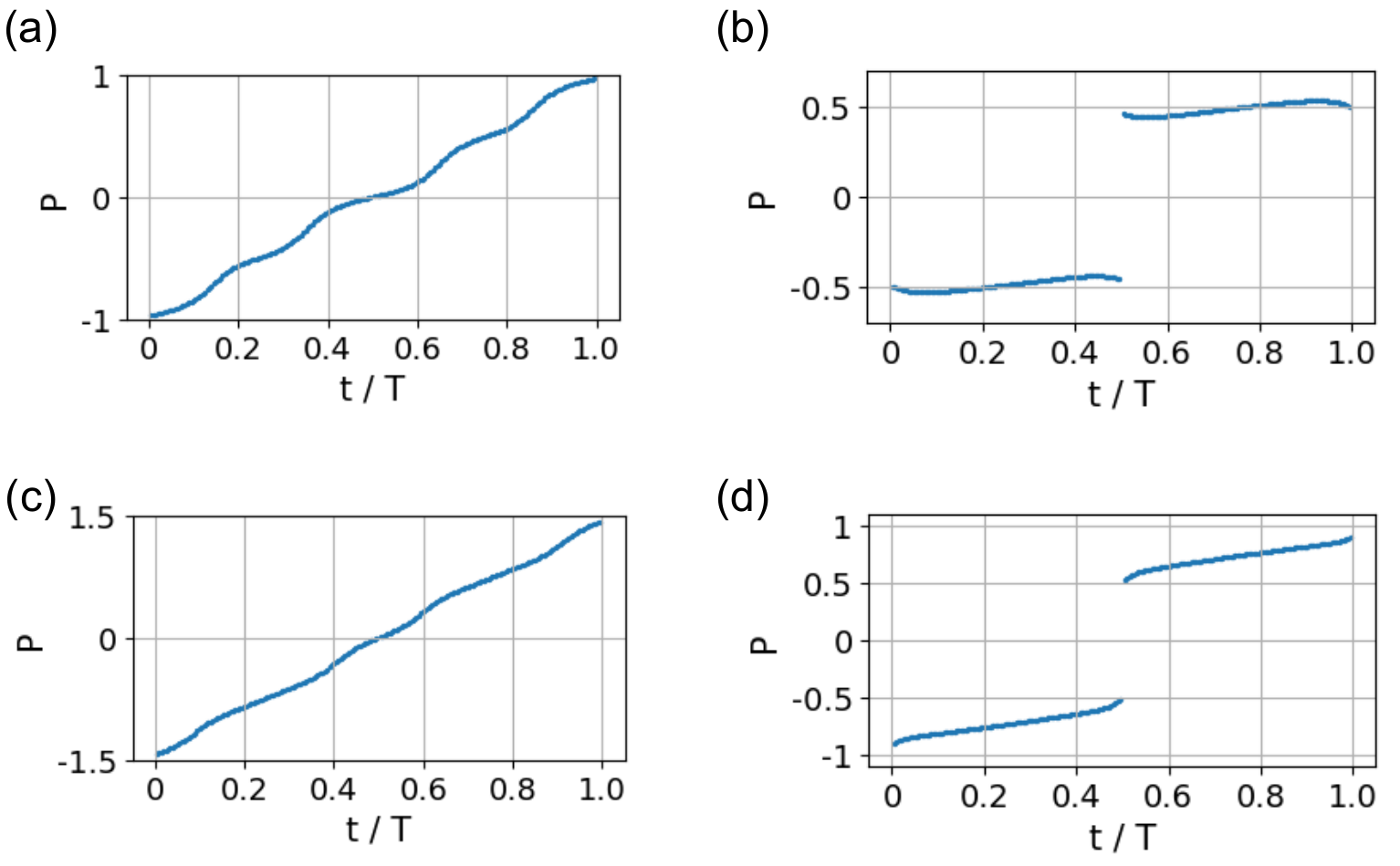}
\caption{The sCoM as a function of $t/T$ for
(a) $N=2, \delta_0=0.8$,
(b) $N=2, \delta_0=0.2$,
(c) $N=3, \delta_0=0.8$,
and
(d) $N=3, \delta_0=0.2$. 
The sCoM jump per cycle is (a) $I=2$, (b) $I=0$, (c) $I=3$, and {(d)} $I=1$.
We set $L=100$ for $N=2$ and $L=120$ for $N=3$. 
The amplitude of the magnetic field is set $\Delta_0=0.1$ for all cases.}
\label{fig:scom} 
\end{figure}

To confirm the bulk-edge correspondence, we numerically calculate the snap-shot expectation value of the sCoM $P(t)$ by the DMRG method by TeNPy \cite{tenpy,Hauschild2024}.
The behavior of the sCoM for some parameter sets is shown in Fig.~\ref{fig:scom}(a)-\ref{fig:scom}(d). 
For each case, we confirm that the sum of the jump $I$ is consistent with the Chern number in the periodic system obtained in the previous section.
This implies that the bulk-edge correspondence holds.
\par
In particular, we discuss the topological pump with $I=N$ for the $N\textrm{-}$leg ladder from the viewpoint of the edge states.
The pump with $I=N$ can be understood by considering the decoupled limit $\delta_0=J$. 
At $t/T=0$, due to the diagonal edge, the $N$-free spins at each end of the ladder appear resulting in the discontinuous jump of the sCoM as $I =- (-\frac{N}{2}-\frac{N}{2})=N$ induced by the introduction of the magnetic field $H_{\rm SSM}(t=\pm 0)$.
Due to the topological stability, the jump of the sCoM $I$ remains the same when the modulation amplitude $\delta_0$ changes from the limit unless it crosses a critical point of the periodic case,
as shown in Fig.~\ref{fig:scom}(a) and \ref{fig:scom}(c).
\par
We then expect that the discontinuity of the sCoM at $t/T=0$ in Fig.~\ref{fig:scom}(a) and \ref{fig:scom}(c) is due to the spherical effective spin $S_{\mathrm{eff}}=\frac{N}{2}$ based on the Marshall-Lieb-Mattis (MLM) theorem~\cite{LM1962,Marshall}.
To be more specific, according to the MLM theorem, the sign of the correlation function between two spins in the ground state is positive when they are in the same sublattice in a bipartite lattice \cite{Tasaki_text}.
We can apply this to our model of $H_{\mathrm{O}}(t=0)$.
As shown in Fig.~\ref{fig:edge}, the edge spins at each end of $H_{\mathrm{O}}(t=0)$ are in the same sublattice since the shape of the edge is diagonal and the correlation function of two of them is positive.
Then, we expect that the effective couplings of these edge spins are ferromagnetic and the effective edge spin is $S_{\mathrm{eff}} = \frac{N}{2}$, resulting in the discontinuous jump of the sCoM at $t/T=0$ as $I =- (-S_{\mathrm{eff}}-S_{\mathrm{eff}})=N$.

\section{\label{summary} Summary}
We have constructed a topological pump of the $N$-leg spin ladder by introducing the modulation of $2N\textrm{-}$site clusterization.
As for the ground state of the model rich SPT phases appear protected by the time-reversal symmetry.
We have found $N$ critical points by using the quantized Berry phase. 
By adding the staggered magnetic field, we realize the topological pump by extending the parameter space. 
The pump is topologically non-trivial since the pump path encloses the critical points as topological obstructions.
The topological pump with the periodic boundary condition is characterized by the Chern number.
We have also found the plateau transitions of the Chern number by deforming the pump path. 
Then, we have further considered the open boundary case where edge states on the diagonal edges appear and contribute to the topological pump, that is, their creation and annihilation near the edges are complementary to the bulk pump.
The edge states induce a degeneracy.
It implies a singular behavior to the sCoM as the discontinuous jump.
We have also numerically verified the bulk-edge correspondence, namely, the amount of the jump per cycle is equal to the bulk Chern number.

\acknowledgements
This work is supported by 
JST SPRING, Grant No.~JPMJSP2124,
JST-CREST Grant No.~JPMJCR19T1, 
and JSPS KAKENHI, Grant 
No.~23K13026 (YK), JP23K03243 (TM), JP24K22848 (KS), JP23K25788 (YH).

\appendix
\section*{\label{appendix} Appendix: The bulk-edge correspondence of the topological pump}
In this Appendix, we review the bulk-edge correspondence of the topological pump based on Ref.~\cite{hatsugai_fukui}. 
It shows that the pumped spin per cycle described by the Chern number in the twisted boundary condition corresponds to the sum of the jumps of the sCoM in the open boundary condition.
\par
For the derivation, we start with a pump Hamiltonian with the global twist in the periodic boundary condition as
\begin{eqnarray}
    \bar{H}_{\rm P}(\theta,t) = \bar{H}(\theta,t) + H_{\mathrm{SM}}(t),
\end{eqnarray}
whose pump parameters are the same as ${H}_{P}(\theta,t)$ in the main text.
The above pump Hamiltonian is related to ${H}_{\rm P}(\theta,t)$ in Eq.~\eqref{eq:pump_Hamiltonian}
through the large gauge transformation \eqref{eq:large_gauge_transformation} as ${H}_{\rm P}(\theta,t) = U^{\dagger}(\theta) \bar{H}_{\rm P}(\theta,t) U(\theta)$. 
The global twist of $\bar{H}_{\rm P}(\theta,t)$ is gauged out by the large gauge transformation except for the twist on the boundary bonds.
\par
Next, we define another pump Hamiltonian with global twist in the open boundary condition $\bar{H}_{\rm O}(\theta,t)$ by cutting the boundary bonds in $\bar{H}_{\rm P}(\theta,t)$.
For the Hamiltonian $\bar{H}_{\rm O}(\theta,t)$, we can gauge out the global twist by $U(\theta)$ as \cite{hatsugai_fukui} 
\begin{eqnarray}
    {H}_{O}(t) = U^{\dagger}(\theta) \bar{H}_{O}(\theta,t) U(\theta),
\end{eqnarray}
where ${H}_{O}(t)$ is given in the main text in Eq.~(\ref{eq:H_o}) and is $\theta\textrm{-}$independent.
The current operator of $\bar{H}_{O}(\theta,t)$ is given as $\partial_{\theta}\bar{H}_{O}(\theta,t)$ (here, we set $\hbar=1$). 
In the adiabatic approximation, the expectation value of the current is given as
\begin{eqnarray}
    \bar{j}_{\mathrm{O}}(\theta,t) = -i\bar{B}_{\mathrm{O}}(\theta,t) ,
\end{eqnarray}
where $\bar{B}_{\mathrm{O}}(\theta,t)$ is the Berry curvature of $\bar{H}_{O}(\theta,t)$ written as $\bar{B}^{\mathrm{(O)}}(\theta,t) = \partial_{\theta}\bar{A}_{t}^{\mathrm{(O)}}(\theta,t) - \partial_{t}\bar{A}_{\theta}^{\mathrm{(O)}}(\theta,t)$.
$\bar{A}_{\alpha}^{\mathrm{(O)}}(\theta,t)$ is the Berry connection of $\bar{H}_{O}(\theta,t)$ defined as $\bar{A}^{\mathrm{(O)}}(\theta,t) = \langle \bar{g}^{(\mathrm{O})}(\theta,t)| \partial_{\alpha}\bar{g}^{(\mathrm{O})}(\theta,t)\rangle$, where $|\bar{g}^{(\mathrm{O})}(\theta,t)\rangle$ is the  ground state of $\bar{H}_{O}(\theta,t)$.
Here, we take the temporal gauge by further introducing a gauge transformation, so that the gauge field {$\bar{A}_{\alpha}^{(O),(t)}(\theta,t)$} can satisfy $\bar{A}_{t}^{(O),(t)}(\theta,t)=0$.
The expectation value of the current in the open boundary system is also written as
\begin{eqnarray}
    \bar{j}_{\mathrm{O}}(\theta,t) =&& {i\partial_{t}\bar{A}_{\theta}^{\mathrm{(O)},(t)}(\theta,t)},
    \end{eqnarray}
where {$\bar{A}_{\theta}^{\mathrm{(O)},(t)}(\theta,t)\equiv$ $\bar{A}^{\mathrm{(O)}}_{\theta}(\theta,t)-\bar{A}^{\mathrm{(O)}}_{\theta}(\theta,0)-\partial_{\theta}\int_{0}^{t}d\tau\bar{A}^{\mathrm{(O)}}_{t}(\theta,\tau)$}.
\par
The above discussion applies to the pump Hamiltonian $\bar{H}_{P}(\theta,t)$ and the expectation value of the current operator $\partial_{\theta}\bar{H}_{\mathrm{P}}(\theta,t)$ is given as
\begin{eqnarray}
    \bar{j}_{\mathrm{P}}(\theta,t) =&& i\partial_{t}\bar{A}_{\theta}^{\mathrm{(P),(t)}}(\theta,t),
\end{eqnarray}
where {$\bar{A}_{\alpha}^{\mathrm{(P),(t)}}(\theta,t)$ is the Berry connection} and $|\bar{g}^{(\mathrm{P})}(\theta,t)\rangle$ is the snap-shot ground state of $\bar{H}_{P}(\theta,t)$. 
\par
By the large gauge transformation {$U(\theta)$}, the ground states of $\bar{H}_{\mathrm{O}}(\theta,t)$ is mapped as
\begin{eqnarray}
|\bar{g}^{(\mathrm{O})}(\theta,t)\rangle = U(\theta)|{g}(t)\rangle,
\end{eqnarray}
where $|{g}(t)\rangle$ is the ground state of ${H}_{O}(t)$ and is $\theta\textrm{-}$independent.
Then, the current {$\bar{j}_{\mathrm{O}}(\theta,t)$} is written as
{
\begin{eqnarray}
\bar{j}_{\mathrm{O}}(\theta,t) 
     &=& i\partial_{t}[\bar{A}^{\mathrm{(O)}}_{\theta}(\theta,t)-\bar{A}^{\mathrm{(O)}}_{\theta}(\theta,0)-\partial_{\theta}\int_{0}^{t}d\tau\bar{A}^{\mathrm{(O)}}_{t}(\theta,\tau)]\nonumber\\
    &=&i\partial_{t}[\bar{A}^{\mathrm{(O)}}_{\theta}(\theta,t)-\bar{A}^{\mathrm{(O)}}_{\theta}(\theta,0)]\nonumber\\
    &=&i\partial_{t}[-iP(t)+iP(0)]=\partial_{t}P(t),
\end{eqnarray}
}where we used the fact that ${\bar A}^{\mathrm{(O)}}_{t}(\theta,\tau)$ is $\theta$-independent from the first to the second line and $P(t)$ is the sCoM given in the main text.
The total pumped spin per cycle in $\bar{H}_{O}(\theta,t)$ is the time integration of the current, which is written as
\begin{eqnarray}
\bar{Q}^{\mathrm{(O)}}
=&& \int_{0}^{T}dt \bar{j}_{\mathrm{O}}(\theta,t)= \int_{0}^{T}dt \partial_{t}P(t) \nonumber\\
=&& \sum_{i}\int_{t_{i-1}}^{t_{i}}dt \partial_{t}P(t) + \sum_{i} P(t)|_{t_{i}-0}^{t_{i}+0}.
\end{eqnarray}
Due to the time periodicity of the pump, the total pumped spin in the open boundary condition is zero.
We now assume that the pumped spin consists of the contribution from the bulk denoted by $Q^{\mathrm{(O)}}_{\mathrm{bulk}}$ and from the edge denoted by $I$. Their sum cancels out since the total pumped spin is zero:
{
\begin{eqnarray}
Q^{\mathrm{(O)}}_{\mathrm{bulk}} = I,
\end{eqnarray}
where $Q^{\mathrm{(O)}}_{\mathrm{bulk}} = \sum_{i}\int_{t_{i-1}}^{t_{i}}dt \partial_{t}P(t)$ and $I =-\sum_{i} P(t)|_{t_{i}-0}^{t_{i}+0}$.}
\par
Next, we move to the pump with the periodic boundary condition.
In the large system size limit, 
{one can assume} that the $\theta\textrm{-}$average {of the} total pumped spin per cycle of the pump with the periodic boundary condition, denoted by ${\bar{Q}^{\mathrm{P}}_{\mathrm{bulk}}(\theta)}$, is equal to that in the open boundary condition as
\begin{eqnarray}
    Q^{\mathrm{(O)}}_{\mathrm{bulk}} = \frac{1}{2\pi}\int_{0}^{2\pi}d\theta \bar{Q}^{\mathrm{(P)}}_{\mathrm{bulk}}{(\theta)}.\nonumber
\end{eqnarray}
We substitute $\bar{Q}^{\mathrm{(P)}}_{\mathrm{bulk}}{(\theta)}=\int_{0}^{T}dt \bar{j}_{\mathrm{P}}(\theta,t)$ and write the pumped spin as
\begin{eqnarray}
    Q^{\mathrm{(O)}}_{\mathrm{bulk}} =&& \frac{1}{2\pi}\int_{0}^{2\pi}d\theta \int_{0}^{T}dt \bar{j}_{\mathrm{P}}(\theta,t)\nonumber\\
    =&& \frac{1}{2\pi}\int_{0}^{2\pi}d\theta \int_{0}^{T}dt i \partial_{t}\bar{A}_{\theta}^{\mathrm{(P),(t)}}(\theta,t)\nonumber\\
    =&& \frac{1}{2\pi i}\int_{0}^{2\pi}d\theta \int_{0}^{T}dt [-\partial_{t}\bar{A}_{\theta}^{\mathrm{(P),(t)}}(\theta,t)]\nonumber\\
    =&& \frac{1}{2\pi i}\int_{0}^{2\pi}d\theta \int_{0}^{T}dt \bar{B}^{(\mathrm{P})}(\theta,t)\nonumber\\
    =&& \bar{C},
\end{eqnarray}
where $\bar{B}^{(\mathrm{P})}(\theta,t)$ is the Berry curvature and $\bar{C}$ is the Chern number of $\bar{H}_{\mathrm{P}}(\theta,t)$.
{Then, we obtain the following:}
\begin{eqnarray}
    I = \bar{C}. \label{eq:i_cbar}
\end{eqnarray}
\par
In fact, the Chern number {$\bar{C}$} of the pump Hamiltonian $\bar{H}_{\mathrm{P}}(\theta,t)$ is equal to that of ${H}_{\mathrm{P}}(\theta,t)$ in the main text.
To see this, we first point out that the Berry curvatures are mapped as
\begin{eqnarray}
    \bar{B}^{\mathrm{(P)}}(\theta,t) =&& B^{(\mathrm{P})}(\theta,t)-i\partial_{t} \langle g(\theta,t)|U^{\dagger}(\theta)\partial_{\theta}U(\theta)|g(\theta,t)\rangle  \nonumber\\
    =&& B^{(\mathrm{P})}(\theta,t)+\partial_{t} \langle g(\theta,t)|\sum_{j=0}^{L-1}\frac{j-j_{\mathrm{c}}}{L}S_{j}^{z}|g(\theta,t)\rangle,\nonumber\\
    \label{eq:berry_curv}
\end{eqnarray}
where {$|g(\theta,t)\rangle$} is the ground state of ${H}_{\mathrm{P}}(\theta,t)$ in the main text.
This is because of the relation, ${H}_{\mathrm{P}}(\theta,t) = U^{\dagger}(\theta) \bar{H}_{\mathrm{P}}(\theta,t) U(\theta)$.
Then the Chern number $\bar{C}$ is
\begin{eqnarray}
    \bar{C} =&& \frac{1}{2\pi i}\int_{0}^{2\pi}d\theta \int_{0}^{T}dt \biggr[ {B}^{(\mathrm{P})}(\theta,t)\nonumber\\
    && +\partial_{t} \langle g(\theta,t)|\sum_{j=0}^{L-1}\frac{j-j_{\mathrm{c}}}{L}S_{j}^{z}|g(\theta,t)\rangle \biggr] \nonumber\\
    =&& \frac{1}{2\pi i}\int_{0}^{2\pi}d\theta \int_{0}^{T}dt {B}^{(\mathrm{P})}(\theta,t)\nonumber\\
    =&& C,\label{eq:c_cbar}
\end{eqnarray}
where $C$ is the Chern number for ${H}_{\mathrm{P}}(\theta,t)$.
Note that the second term of Eq.~(\ref{eq:berry_curv}) does not contribute to the Chern number.
Combining Eqs.~(\ref{eq:i_cbar}) and (\ref{eq:c_cbar}), we find that the bulk-edge correspondence $C=I$ in the main text is established.

\bibliographystyle{apsrev4-2}
\bibliography{ref}

\begin{thebibliography}{56}
\bibitem [{cal()}]{cal_BP}%
  \BibitemOpen
  \href@noop {} {\ }\bibinfo {note} {{The Berry phase is calculated as follows
  \begin{eqnarray} &&-i\int^{2\pi}_{0}d\theta \langle
  g(\theta)|V(\theta)\partial_{\theta}(V^{\dagger}(\theta)|g(\theta)\rangle)\nonumber\\
  &&=\gamma(\delta)-i\int^{2\pi}_{0}d\theta \sum^{L-1}_{j=L-N}[\langle
  g(\theta)|S^{z}_j|g(\theta)\rangle -S]\nonumber\\ &&=\gamma(\delta)+\pi N,
  \nonumber \end{eqnarray} where we used $\langle
  g(\theta)|S^{z}_j|g(\theta)\rangle=0$ due to the time-reversal
  symmetry.}}\BibitemShut {Stop}%
\end{thebibliography}
\end{document}